\documentclass[12pt,a4]{iopart}
\usepackage{iopams}  
\usepackage{epsfig}  

\def\R{\mathbb{R}}

\begin{document}
\paper{Statistical Self-Similarity 
of One-Dimensional Growth Processes}

\author{Michael Pr\"ahofer and Herbert Spohn}

\address{Zentrum Mathematik and Physik Department, TU M\"unchen, 
  D-80290 M\"unchen, Germany. Email: \emph{praehofer@ma.tum.de, 
    spohn@ma.tum.de}}

\begin{abstract}
  For one-dimensional growth processes we consider the distribution of the
  height above a given point of the substrate and study its scale invariance
  in the limit of large times. 
  We argue that for self-similar growth 
  from a single seed the universal distribution is the Tracy-Widom 
  distribution from 
  the theory of random matrices and that for growth from a flat substrate it is
  some other, only numerically determined distribution. In particular, for 
  the polynuclear growth model in the droplet geometry the height maps onto 
  the longest 
  increasing subsequence of a random permutation, from which the 
  height distribution is identified as the Tracy-Widom distribution.
\end{abstract}



\section{Introduction}
Quite analogous to equilibrium systems close to their critical point,
growth processes exhibit statistical self-similarity and scaling. 
Their theoretical investigation is a continuing enterprise for a variety of 
reasons. Perhaps the most fundamental one is that statistical self-similarity 
emerges from the dynamical rules without any particular fine-tuning, which 
is believed to be a basic characteristic of many nonequilibrium systems 
\cite{BS95,B96,F95}. The persisting theoretical  challenge is to 
develop renormalization group type methods which have been so powerful in the 
understanding of equilibrium critical phenomena.

In the context of KPZ theory \cite{KPZ86} most of the recent efforts 
have concentrated on scaling exponents and
in particular on the question of an upper critical dimension 
\cite{LK97,NS97,L98,CMMP99}. In this note we will focus on the full height 
distribution at one point above the substrate. In addition we will restrict 
ourselves to one-dimensional growth processes, where {\em static} fluctuations
are the same as in equilibrium due to a fluctuation-dissipation property,
i.e. the roughness exponent is known to be $\frac12$.

The most commonly adopted set-up is to start from a flat substrate, i.e., if
$h(x,t)$ denotes the height above $x\in\R$ at time $t$, then initially 
$h(x,0)=ux$ with slope $u$. $h(x,t)$, $t>0$, is random through the dynamical 
rules which will have to be specified for a concrete model. The slope $u$ 
is a locally conserved quantity.
By translation invariance, $h(x,t)-ux$ has a distribution independent of $x$ 
and we may consider the height $H(t)=h(0,t)$ above the origin. 
For large times it will increase linearly with time with a slope dependent 
growth velocity $v(u)$.~$H(t)$ fluctuates on the 
scale $t^{1/3}$. Therefore 
\begin{equation}
  \label{eq:flatscalingform}
  H(t)=v(u)t+C(u)t^{1/3}\xi
\end{equation}
for large $t$. The random variable $\xi$ is of order one and 
independent of $t$ 
and $u$. Equation (\ref{eq:flatscalingform}) should be viewed in analogy 
to the central limit theorem for a random walk with independent 
increments. In that case $v(u)$ would be the average drift, the 
fluctuations diffusive on the scale $t^{1/2}$, and $\xi$ 
 a standard Gaussian random variable.

Following Krug {\em et al} \cite{KMH92}, the prefactor $C(u)$ depends only on 
macroscopic properties of the growth model, namely the leading nonlinear part
of the slope dependent growth velocity, $\lambda(u)=v''(u)$, 
and the static susceptibility or roughness amplitude,
\begin{equation}
  \label{eq:roughnessamplitude}
  A(u)=\lim_{x\to\infty}\lim_{t\to\infty}x^{-1}
  \langle(h(x,t)-h(0,t)-ux)^2\rangle,
\end{equation}
with limits in this order. On dimensional grounds we must have then
\begin{equation}
  \label{eq:coefficient}
  C(u)=\mbox{sign}(\lambda(u))\left(a|\lambda(u)|A(u)^2\right)^{1/3},
\end{equation}
where $a$ is a  dimensionless number, independent of $u$, fixing the
absolute scale of $\xi$. 
Krug {\em et al} \cite{KMH92} 
have studied $\xi$ 
on the level of moments for a variety  of growth models, thereby strongly
supporting the claim that the scaling form (\ref{eq:flatscalingform}) 
together with
(\ref{eq:coefficient}) is the same for all growth models in the KPZ
universality class. 

In the very early studies on growth models, like the Eden model, one started 
with a single seed at the origin to have a cluster growing through 
irreversible 
aggregation. For large times the droplet grows then self-similarly with a 
shape depending on the particular dynamical rule. In fact the macroscopic 
shape 
is determined through the slope dependent growth velocity by the Wulff 
construction, i.e. growth velocity  and shape  are related as 
surface tension and crystal shape in equilibrium.
Thus, as for a flat substrate, in the droplet geometry there is 
self-similar growth, but now with a (macroscopically) non-vanishing
curvature. The analogue of (\ref{eq:flatscalingform}) is the distance of
the surface from the origin along a given ray and it is natural to expect
for it the scaling form
\begin{equation}
  \label{eq:dropletscalingform}
  R(t)=vt+Ct^{1/3}\chi
\end{equation}
with some universal random variable $\chi$.

The main part of our note studies the shape fluctuations 
(\ref{eq:dropletscalingform}) in the droplet geometry. In Section 
\ref{sec:droplet} we show that for the polynuclear growth (PNG) model
(see for example \cite{BBDK98} and references therein)
 $\chi$ has the 
Tracy-Widom (TW) distribution appearing in the theory of random matrices. 
We give some numerical and analytical support for the universality
hypothesis. In addition, we clearly demonstrate that $\xi\neq\chi$.
Thus for self-similar growth with zero noise initial conditions
there are two distinct classes, flat substrate
(zero curvature) and droplet (non-zero curvature). In both geometries
the height has the scaling form (\ref{eq:flatscalingform}), resp.
(\ref{eq:dropletscalingform}), but the universal random variables
$\xi$ and $\chi$ have different distributions.

\section{The PNG droplet}
\label{sec:droplet}
The PNG model describes a crystal growing
layer by layer on a one-dimensional substrate through random deposition of 
particles. They nucleate on existing plateaus of the crystal forming new 
islands. In an idealization these islands spread 
laterally with constant speed by steady condensation of further
material at the edges of the islands. Adjacent islands of the same level 
coalesce upon meeting and on top of the new levels further islands emerge.

Nucleation events occur independently and uniformly in space-time. 
Given the space-time coordinates $(x_0,t_0)$, $x_0\in\R$, 
$t_0>0$, of such an event, the corresponding island nucleates in level 
\begin{equation}
\label{eq:whichlevel}
h_0=h(x_0,t_0)+1
\end{equation}
in units of transverse lattice spacings. We also choose the time unit such that
the lateral growth speed equals one. 
To determine the height $h(x,t)$ we note that it depends only on the 
nucleation events in the backward light cone of $(x,t)$. We label them as 
$(x_n,t_n)$, $n=1,2,\ldots$. Then 
\begin{equation}
\label{eq:detlevel}
h_n=h(x_n,t_n)+1,\quad n=1,2,\ldots,
\end{equation}
according to (\ref{eq:whichlevel}), and $h(x,t)$ is obtained as the maximum 
over all $h_n$ in the backward light cone,
\begin{equation}
\label{eq:height}
h(x,t)=\max\{h_n:n=1,2,\ldots,\,|x-x_n|<t-t_n\},
\end{equation}
with the rule  that $h(x,t)=0$ if there is no such nucleation event.
(\ref{eq:detlevel}) and (\ref{eq:height}) together define recursively all 
$h_n$ and therefore $h(x,t)$.

For droplet growth on the initially flat substrate, a single island starts 
spreading from the origin and further nucleations 
take place only above this ground layer which we define to have zero height.
We want to study the probability distribution of $h(x,t)$, 
i.e.~$\mbox{Prob}\{h(x,t)=n\}$ for $n=0,1,2,\ldots$, $|x|<t$.
Clearly $h(x,t)$ is determined by the set of nucleation events 
inside
the rectangle $R_{(x,t)}=
\{(x',t')\in\R^2:\,\,|x'|\leq t'\mbox{ and }|x-x'|\leq t-t'\}$.
In light-like coordinates, $u=(t'+x')/\sqrt{2}$, $v=(t'-x')/\sqrt{2}$, the 
rectangle $R_{(x,t)}$ equals $[0,U]\times[0,V]$  with $U=(t+x)/\sqrt{2}$,
$V=(t-x)/\sqrt{2}$. The nucleation events in $R_{(x,t)}$ are Poisson 
distributed with density one.
We label them as $(u_n,v_n)$, $n=1,\ldots,N$, $N$ random, such that 
$0\leq u_1<\cdots<u_N\leq U$. The corresponding order in the second 
coordinate $v$, $0\leq v_{p(1)}<\cdots<v_{p(N)}\leq V$,
defines then (with probability one) a permutation $p$ of length $N$, 
compare with Figure 1. 
In order
to distinguish nucleation events corresponding to different height levels,
we partition $(p(1),\ldots,p(N))$ into decreasing subsequences according to
the following algorithm.
For the first subsequence one starts with the first entry in the tuple 
and, scanning from left to right, one adds an entry to the subsequence if 
it is smaller than the so far last entry in this subsequence. Every
used entry is marked as deleted. When the end of the tuple is reached the first
subsequence is completed. This procedure is repeated to obtain further
subsequences by using only the undeleted entries of the original tuple. 
As a result the permutation $p$ is partitioned into decreasing subsequences. 
The nucleation events of each such subsequence are on the same height level
and distinct subsequences correspond to distinct heights. The
space-time height lines, across which $h(x',t')$ increases by one unit, are 
thus constructed by connecting the nucleation events belonging to the same
subsequence by a zigzag line, as depicted in Figure 1.
\begin{figure}[htbp]
  \label{fig:steps}
  \begin{center}
    \mbox{\epsfig{file=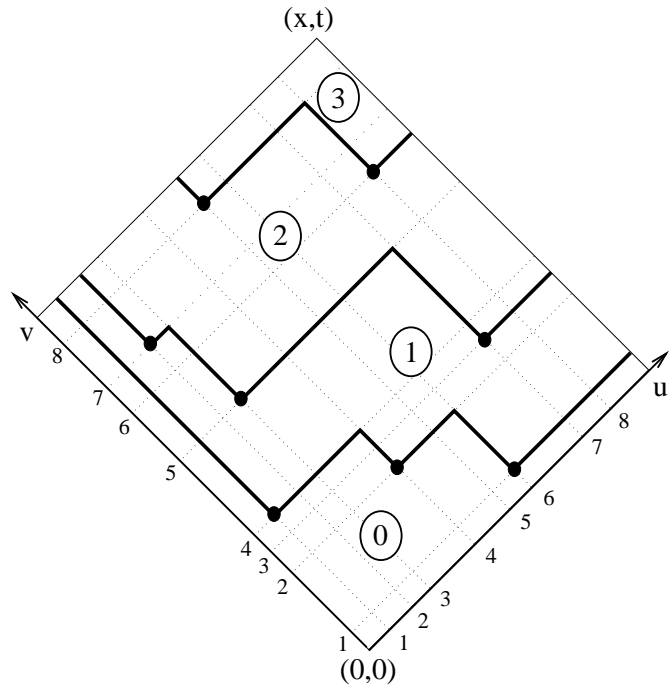,height=9cm}}
  \end{center}
  \caption{Space-time picture of a tiny PNG droplet} 
\end{figure}
In this example we have the permutation $(4,7,5,2,8,1,3,6)$. The
decreasing subsequences are $(4,2,1)$, $(7,5,3)$, $(8,6)$. They define
three height lines. For an arbitrary permutation $p$, $h(x,t)$ equals the 
number of 
decreasing subsequences, which is precisely the length, $l$, of the {\em
  longest increasing subsequence} of $p$.  To see this take any
increasing subsequence of $p$. By construction every element of this 
subsequence belongs to a different height line. Therefore $l\leq h(x,t)$. 
The reverse inequality is obtained by noticing that a nucleation event 
$(i,p(i))$ at level $k+1$ is necessarily
situated in the forward light cone of at least one nucleation event 
$(j,p(j))$ at level $k$,
i.e. $i>j$ and $p(i)>p(j)$. $(j,p(j))$ is called a predecessor of $(i,p(i))$. 
Since there is at least one nucleation event in level 
$h(x,t)$, successive selections of predecessors in consecutive levels result
in an increasing subsequence of length $h(x,t)$, therefore $l\geq h(x,t)$.

In fact, the mapping described here has been used before in \cite{AD95}, 
where the authors consider the Hammersley process \cite{H72} in order to 
study the asymptotics of $l$ as $N\to\infty$.
We remark that the world lines of particles in 
the Hammersley process are exactly the space-time step lines of the
PNG model, after a rotation of $45^\circ$ together with an appropriate 
adjustment of the boundary conditions. 
To our knowledge this simple observation has not yet been reported in the 
literature and establishes a 
connection between the two models in much the same way as between 
the simple exclusion and the zero-range processes \cite{L85}.

For given $N$ the nucleation events are uniformly and independently distributed
in the rectangle $R_{(x,t)}$. Since  the $u$-- and $v$--integrations 
factorize, this induces the uniform distribution on the set of all 
permutations of length $N$. By the Poisson law, $N$ has the probability 
distribution
\begin{equation}
\label{eq:poisson}
P(N)=\exp(-\lambda)\frac{\lambda^N}{N!},\mbox{ $\lambda=UV=(t^2-x^2)/2$, 
$N=0,1,\ldots$.}
\end{equation}
Thus the height of the PNG droplet is equal to the length of the longest 
increasing subsequence of a random permutation with Poisson distributed 
length, which for a fixed length of the permutation is known as Ulam's 
problem \cite{U61}.

In recent years  there has been spectacular
advance on Ulam's problem, for a survey see \cite{AD99}. One important
element are deep combinatorial identities 
expressing the probability distribution of $l$ through quantities known
from the theory of random matrices. In our particular context the identity 
reads
\begin{equation}
\label{eq:finite}
\hspace{-2cm}\mbox{Prob}\{l(\lambda)=n\}=\frac{\exp(-\lambda)}{n!(2\pi)^n}
\int_{[-\pi,\pi]^n}\hspace{-5mm}
\exp\Big(2\sqrt{\lambda}\sum_{j=1}^n\cos\theta_j\Big)
\prod_{1\leq j<k\leq n}\hspace{-3mm}|e^{i\theta_j}-e^{i\theta_k}|^2d^n\theta,
\end{equation}
which corresponds to the partition function for the eigenvalues in the 
unitary ensemble with an external cosine potential.
Baik \emph{et al} \cite{BDJ98} succeeded  to extract from (\ref{eq:finite})
the $\lambda\to\infty$ asymptotics with the results
\begin{equation}
\label{eq:limit}
l(\lambda)/\sqrt{\lambda}\to2\quad\mbox{in probability}
\end{equation}
and
\begin{equation}
\label{eq:scaling}
\frac{l(\lambda)-2\sqrt{\lambda}}{\lambda^{1/6}}\to\chi \mbox{ in 
distribution},
\end{equation}
together with the convergence of moments. Here $\chi$ is a 
random variable, whose distribution is the famous Tracy-Widom (TW) distribution
of the properly shifted and rescaled largest eigenvalue of a GUE distributed
random matrix \cite{TW94}. $\chi$ has the  distribution function 
\begin{equation}
\label{eq:TW}
F(t)=\exp\bigg(-\int_t^\infty(x-t)u(x)^2dx\bigg),
\end{equation}
where $u(x)$ is the unique solution of the Painlev\'e II equation,
\begin{equation}
\label{eq:painleve}
 u''=2u^3+xu,
\end{equation}
with the asymptotic boundary conditions $u(x)\sim\mbox{Ai}(x)$ 
as $x\to\infty$, $\mbox{Ai}(x)$ being the Airy function \cite{HM80}.
The probability density, $F'(t)$, of $\xi$ has the asymptotics 
$\ln F'(t)\simeq-|t|^3/12$ for $t\to-\infty$ and 
$\ln F'(t)\simeq-\frac43t^{3/2}$ for $t\to\infty$.

We remark that the connection of the right-hand side of equation 
(\ref{eq:finite}) to the Painlev\'e II equation has been established on an
heuristic level already in the context of unitary-matrix models \cite{PS90}.

To come back to the PNG droplet we only have to use that $h(x,t)=l(\lambda)$
and to translate (\ref{eq:limit}) and (\ref{eq:scaling}) into the surface 
picture. Equation (\ref{eq:limit}) states that the droplet acquires for 
large times a deterministic ellipsoidal shape given by
\begin{equation}
\label{eq:macroscopic}
\lim_{t\to\infty}t^{-1}h(ct,t)=\sqrt{2}\sqrt{1-c^2}.
\end{equation}
for $-1\leq c\leq 1$. Thus with our particular initial conditions the PNG 
model grows in the 
form of a droplet. The second result ensures that the shape fluctuations 
are on the scale $t^{1/3}$
and in addition identifies the limiting distribution,
\begin{equation}
\label{eq:distribution}
\lim_{t\to\infty}\frac{h(ct,t)-t\sqrt{2}\sqrt{1-c^2}}
{(t\sqrt{(1-c^2)/2})^{1/3}}=\chi. 
\end{equation}
Note that in (\ref{eq:distribution})
only the asymptotic average is subtracted. Therefore $\chi$ may have a 
mean different from zero, which is indeed the case, 
as can be seen in Figure 2,
where the TW distribution $F'(t)$ is plotted on a semi-logarithmic scale.

\section{Flat substrate}
For flat initial conditions, by translational invariance, it suffices to 
study the distribution of the height above the origin, i.e. $H(t)=h(0,t)$.
At first glance, since the droplet curvature vanishes on the microscopic 
scale, one would expect the distributions for $\xi$ in Equation 
(\ref{eq:flatscalingform}) and for $\chi$ in Equation
(\ref{eq:dropletscalingform})
to be identical. To better understand this point we again consider the
PNG model.

The flat initial conditions in the PNG model are $h(x,0)=0$ and 
nucleation is now allowed everywhere on the line.  
We consider $H(t)=h(0,t)$. 
$H(t)$ depends on nucleation events in the backward light cone of $(0,t)$,
i.e. in the triangle $T_t=\{(x',t');\,\,0\leq t'<t$, $|x'|<t\}$. As before we 
construct a random permutation from the nucleation events inside $T_t$ and 
$H(t)$ equals the length of the longest increasing subsequence. Unfortunately
the induced measure on the permutations is no longer uniform and the analogue 
of (\ref{eq:finite}) does not seem to be known.

To relate $H(t)$ to the distribution $h(0,t)$ of the height process from the 
PNG droplet we define $H_x(t)$, $|x|\leq t$ as the height
which results if the height lines are constructed from  nucleation events 
inside the rectangle $\{(x',t'):\,\,|x'|\leq t-t'\mbox{ and }|x-x'|\leq t'\}$ 
only. Since a longest increasing subsequence of nucleation events for $H(t)$
must be contained in at least one of these rectangles, we have 
$H(t)=\max_{|x|\leq t}H_x(t)$. Now for a given set of nucleation events in 
$T_t$ the mapping $t'\mapsto t-t'$, $x'\mapsto -x'$ yields a realization
of nucleation events for the PNG droplet up to time $t$. The algorithm to
determine $h(x,t)$, $|x|\leq t$, for the droplet involves the same permutation
as for $H_x(t)$, only the partition has to be into {\em increasing} 
subsequences starting from the {\em right}. Nevertheless $h(x,t)$ equals the 
longest increasing subsequence (from the left) by the same reasoning as in 
Section \ref{sec:droplet}. As a result we have the alternative expression 

\begin{equation}
  \label{eq:max}
H(t)=\max_{|x|\leq t}h(t,x)\quad\mbox{in distribution},
\end{equation}
where $h(x,t)$ is the height process from the PNG droplet.
 
We note that the maximum is taken among highly correlated random variables.
To model their statistics we assume local stationarity and write for the height
of the droplet $h(x,t)=h(0,t)-{x^2}/(t\sqrt{2})+b(x)$ 
in the vicinity of $x=0$,
where $b(x)$ is a two sided random walk with $b(0)=0$, zero mean
and the covariance $\langle b(x)^2\rangle\propto|x|$, as long as $x=o(t)$. 
Therefore the maximum in (\ref{eq:max}) is achieved inside a region of 
order $t^{2/3}$ around $x=0$ and the modifications between the
distributions of $H(t)$ and $h(0,t)$ are on the scale $t^{1/3}$. 
Our argument suggests that $\chi\neq\xi$, which is indeed supported by 
the numerical simulations of Section \ref{sec:numerics}. In addition it
points at a close but non-trivial connection between the
distributions $\xi$ and $\chi$ which might pave the way for an analytical
description of $\xi$.

\section{Universality}
As explained in the Introduction, when starting with a flat substrate of 
slope $u$, the height at the origin scales as 
\begin{equation}
  \label{eq:flatscaling}
    H(t)=v(u)t+C(u)t^{1/3}\xi.
\end{equation}
$v(u)$ and $C(u)$ are non-universal parameters which
depend on the slope $u$ and have to be computed separately for each
model according to the definition in (\ref{eq:coefficient}) and above.
However, $\xi$ is universal under the stated initial conditions.

Clearly it is desirable to have a corresponding scaling form for droplet
growth, the combining feature being the scale invariance of the 
macroscopic shape, which means that for large $l$ we have
\begin{equation}
  \label{eq:detscaling}
  \bar h(lx,lt)=l\bar h(x,t)
\end{equation}
with $\bar h$ denoting the deterministic profile. For a flat substrate 
(\ref{eq:detscaling}) holds trivially, whereas for a droplet the shape
is determined by the Wulff construction (see for example \cite{KS92}). 
In other words, the deterministic growth equation 
\begin{equation}
  \label{eq:deterministic}
  \partial_t \bar h(x,t)
  =v(\partial_x\bar h(x,t))
\end{equation}
with the inclination dependent growth velocity $v$ allows only for two types 
of self-affine solutions,
the flat surface with slope $u$,
$\bar h(x,t)=t\,v(u)+xu$ and the droplet 
\begin{equation}
  \label{eq:dropletmacr}
  h(ct,t)=t\,f(c),
\end{equation}
where $f$ solves the equation $f(c)-cf'(c)=v(f'(c))$.
The parameter $c$ is related to the local slope $u=f'(c)$ by $c=-v'(u)$.
In view of (\ref{eq:flatscaling}) we are therefore led to the 
universal scaling law for the height distribution of a growing droplet
\begin{equation}
  \label{eq:dropletscaling}
  h(ct,t)=tf(c)+\mbox{sign}(\lambda)
  ({\textstyle\frac12}|\lambda|A^2t)^{1/3}\chi.
\end{equation}
Here $\chi$ is a random variable with TW distribution.

With the rigorous result (\ref{eq:distribution})
we are in the position to check the scaling form (\ref{eq:dropletscaling}) 
for the PNG droplet.
In the steady states of the PNG model the left and right island edges 
(steps) have an
independent Poisson distribution whose strength is determined by the average 
slope \cite{BBLT81,KS89}.
With this information one obtains, in our units,
the slope dependent growth velocity
$v(u)=\sqrt{2+u^2}$, yielding $\lambda=2/(2+u^2)^{-2/3}$, 
and the static roughness amplitude $A=v(u)$. 
The relation $c=-u/\sqrt{2+u^2}$ is invertible, therefore one can write
 $\lambda A^2=\sqrt{2(1-c^2)}$. 
Comparing with (\ref{eq:distribution}) we see that
(\ref{eq:dropletscaling}) is indeed satisfied for the PNG droplet.
In addition, with the definition (\ref{eq:TW}) of the TW distribution,
the dimensionless scale factor, denoted by $a$ in Equation 
(\ref{eq:coefficient}), is fixed to be $\frac12$.

In a remarkable paper \cite{J99} Johansson treats the totally asymmetric 
simple exclusion process (TASEP) by a similar but somewhat more involved 
mapping to increasing subsequences in generalized permutations. 
Interpreted as a one-dimensional
discrete growth model it describes a crystal occupying initially three 
quadrants of the plane. Further particles, represented as unit squares, may 
nucleate only if at least two nearest neighbors
are already occupied. Thus the crystal grows into the forth quadrant starting 
from the origin. It has been known for some time \cite{R81} that on a
macroscopic scale the shape becomes deterministic. Johansson proves a
scaling law similar to (\ref{eq:distribution}) with the same limiting
distribution, but a negative prefactor.
In the TASEP particles jump, subject to the exclusion constraint, 
to the right with rate $1$. Since the stationary measures 
are Bernoulli,
one has  $v(u)=\frac14(1-u^2)$, $|u|\leq1$ and $A(u)=\frac14(1-u^2)$. 
Inserting in (\ref{eq:dropletscaling}), we recover the result of Johansson, 
the negative prefactor coming from
$\lambda(u)=-\frac12$. This is one of the few instances 
where universality has been rigorously established on the level of 
distributions.

\section{Numerics}
\label{sec:numerics}
In Figure 2 we compare the two distributions of $\chi$ and $\xi$ in a 
semi-logarithmic plot. The TW distribution, which is drawn as a solid line,
is obtained by numerically solving 
(\ref{eq:painleve}). Its mean is $-1.771087$ and its
variance is $0.813195$. For the skewness,
$\langle\chi^3\rangle_c/\langle\chi^2\rangle_c^{3/2}$, and the kurtosis,
$\langle\chi^4\rangle_c/\langle\chi^2\rangle_c^{2}$, one obtains
 $0.2241$ and $0.09345$, respectively. Here $\langle\chi^k\rangle_c$ denotes 
the $k$-th cumulant (fully truncated moment).
Superimposed are the empirical distributions determined at various times, 
ranging from $1000$ to $2000$ units, from a Monte-Carlo simulation of the 
PNG droplet. Finite time effects are most pronounced for the first moment.
At $t=2000$ there is still a noticeable deviation of about $3\%$ which seems
to fade away very slowly. The other quantities agree with the exact values 
within the statistical fluctuations.

\begin{figure}[htbp]
\label{fig:distributions}
  \begin{center}
    \mbox{\epsfig{file=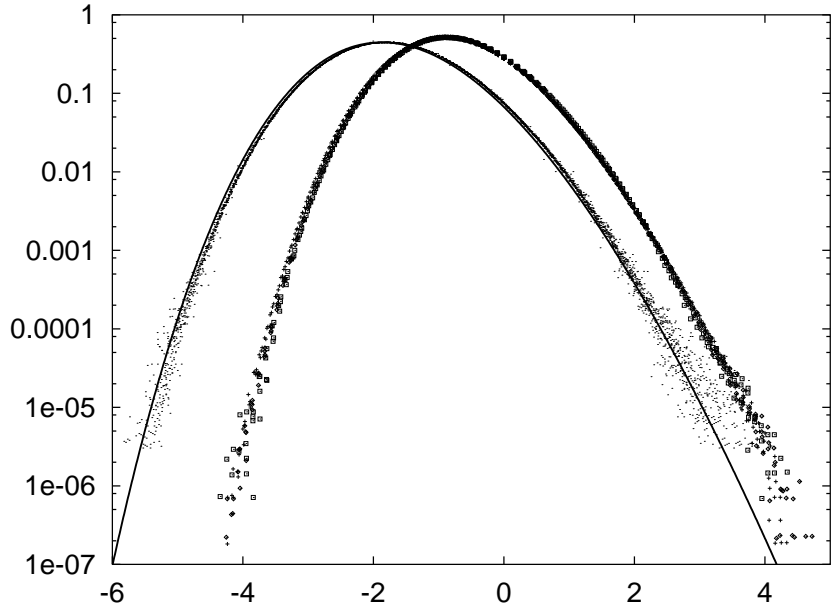,height=8.5cm}}
    \caption{Semi-logarithmic plot of the scaled height distributions. 
      The solid line is the Tracy-Widom distribution. Superimposed are Monte 
      Carlo results for the PNG droplet. The right data points are Monte Carlo
      results for the flat case as explained in the text.}
  \end{center}
\end{figure}

Since for the distribution of $\xi$ there is no analytic expression available,
we performed Monte Carlo simulations of the continuum PNG model for flat 
and equidistantly stepped initial conditions. We used  slopes $0$, $\frac12$ 
and $1$, up to only moderate times $t\leq2500$ on an effectively infinite
substrate.
The number of independent runs
were around $10\,000$ for each slope, but due to translation invariance the 
number of sample points is about $10^7$ for each recorded time. From our data 
we estimate 
the mean, variance, skewness, and kurtosis of $\xi$ as  $-0.76(1)$, 
$0.637(3)$, 
$0.29(1)$, and $0.16(1)$, respectively. In view of the slow convergence of the
first moment for the PNG droplet the estimate for $\langle\xi\rangle$ has to
be taken with care, since it might not yet have reached its true asymptotic 
value.
Our values are consistent with earlier studies of $\xi$ for various growth 
models \cite{KMH92,KMB91}. 

So far we started the growth process with a deterministic height profile.
An obvious variant is to start with a random configuration. The most 
natural choice is to take a stationary height statistics obtained in 
the long time limit. If $h(x,t)$ denotes this stationary growth process (i.e.
the statistics of $\partial_xh(x,t)$ depends only on space-time differences), 
then as before
\begin{equation}
  \label{eq:stationary}
  h(v'(u)t,t)-h(0,0)\simeq v(u)t+C(u)t^{1/3}\eta
\end{equation}
for large $t$. Along a trajectory with slope different from $v'(u)$ one 
has ordinary diffusive scaling as $t^{1/2}$ due to the static roughness
of the height profile.
Our simulations (not shown) yield, the mean being zero by definition,
1.476(7), 0.375(5), and 0.31(1) for the variance, skewness, and kurtosis
of $\eta$, 
which again is consistent with \cite{KMH92} and \cite{KMB91}.

\section{Conclusions}
We have identified three distinct universal distributions: growth with noisy
stationary initial data, and curved and flat growth from deterministic 
initial data.
In addition we have pointed out that the height of the PNG droplet is 
simply related to
the statistics of the length of the longest increasing subsequence of a random
permutation. This leads to a complete characterization of the macroscopic
shape and of the fluctuations in the height above a single reference point. 

It is somewhat unexpected to have the links one-dimensional surface growth
to random permutations to random matrices. As a first introduction to the 
latter we recommend
the survey paper by Aldous and Diaconis \cite{AD99}. Longest increasing
subsequences arise there in
the context of solitaire or patience card games. In patience sorting 
a well shuffled deck of $N$ cards, labeled $1,2,\ldots,N$, is dealt into piles 
according to the, admittedly somewhat boring, two rules: (i) A low card is 
always placed on the leftmost pile with a higher card on top. (ii) If
(i) cannot be achieved, a new pile is started to the right of the previous
piles. The connection
to the PNG model is immediate. A pile of cards corresponds to a 
height line,
the pile size is the number of nucleation events on the height line, and, of
course, the number of piles is the height. 
As regards to  surface growth it would be
of great interest to further exploit the random matrix theory in the 
computation of universal properties and to extend this technique to higher 
dimensions.

\ack 
We thank Craig Tracy for providing us with the 
Mathematica$^{\mbox{\tiny\textregistered}}$
code for the TW distribution.
\vspace{4mm}


\section*{Addendum}
In the meantime we have noted that the scaling distribution $\xi$ for a 
flat substrate is the limiting distribution of the largest eigenvalue of 
a GOE distributed random matrix
\cite{TW96}. More precicely, the distribution function of $\xi$ is given by
\begin{equation}
  \label{eq:xidistr}
  F(2^{2/3}t)=\exp\bigg(-\frac12\int_t^\infty(x-t)u(x)^2dx-
  \frac12\int_t^\infty u(x)dx\bigg). 
\end{equation}
Its first four cumulants are $-0.76007$, $0.63805$, $0.2935$, and $0.1652$.
Our proof uses the symmetrized random permutations studied by Baik and Rains 
\cite{BR99}. We are grateful to Percy Deift and Craig Tracy for pointing out
this reference to us.

\end{document}